\documentclass[11pt]{article}
\usepackage{amsmath,amssymb,graphicx,cite,color,url,hyperref,setspace}
\def\baselinestretch{1.15}
\begin{document}

\begin{flushright}
TIFR/TH/22-22 
\end{flushright}

\begin{center}
{\LARGE\bf Indian Contributions to LHC Theory}

{\sl Sreerup Raychaudhuri}

{\small Department of Theoretical Physics, Tata Institute of Fundamental Research}, \\ 
{\small Homi Bhabha Road, Mumbai, 400005, India} \\ 
Email: {\sf sreerup@theory.tifr.res.in} 

{\small Published in: {\it Eur. Phys. J. Spec. Top.} (March 9, 2023) \\
{\sf https://doi.org/10.1140/epjs/s11734-022-00736-x}} 

\end{center}

\def\baselinestretch{1.05}
\begin{quotation}
\small{\noindent Indian scientists began to work on the theoretical aspects of LHC physics from the early 
1980s, at the same time when the rest of the world started taking interest in this then-futuristic topic. 
From this point grew a whole school of collider phenomenologists, who now form a significant fraction of 
the Indian high-energy physics community. This article briefly reviews the growth and contributions of the 
Indian school, on the way describing some of the physics ideas while placing the work in the international 
context, and proceeding thus, brings the story up to date in mid-2022.}
\end{quotation}
\def\baselinestretch{1.15}
\centerline{\footnotesize\sl Dedicated to the memory of the late D.P.~Roy (1941-2017), the father of 
Indian collider theory.}

\hrulefill

\section{The expectant 80s}\label{sec1}

The LHC owes its origin \cite{1} to CERN's far-sighted leaders, L\'eon van Hove, who served from 1976-1980 
as the Scientific Director-General, and Technical Director-General Sir John Adams (1976-1981), who 
resolved that it would be feasible and scientifically rewarding to build the next-generation storage ring 
at CERN. Thus, in 1977, Adams wrote the crucial concept note that the LEP collider\footnote{The Large 
Electron-Positron collider.} tunnel, which would be excavated in the 1980s, should be made large enough to 
accommodate the superconducting magnets required for a future phase of proton acceleration at higher 
energies. This 1977 note may be thought of as the genesis of the LHC. By 1978, it was already well known 
in the community, even though the formal approval for the LEP came only in 1981, and the machine itself 
started running in 1989 \cite{2}. However, the fact that the 20~TeV LHC would eventually come and that the 
physics at such a machine should be studied, was in the air right from the early 1980s onward. Moreover, 
the USA was supposed to build a 40~TeV Superconducting Super Collider (SSC) which, sadly, never 
materialised\footnote{It was sanctioned in 1989, and eventually cancelled in 1993.}.

The 1980s were, in fact, a period of great optimism for high-energy physics --- or, as it was then more 
commonly known, elementary particle physics. One trigger for this may have lain in the decision of the 
Nobel Prize committee, in 1979, to recognise the work of Glashow (1961) \cite{3}, of Salam (1966-1968) 
\cite{4} and of Weinberg (1967) \cite{5} in developing a working model of electroweak unification. 
Although the Higgs boson \cite{6,7,8}, the cornerstone of the theory developed by Weinberg and Salam, 
would be discovered only 3 decades later \cite{9, 10}, the Nobel Prize committee made an unusual departure 
in deviating from its usual conservative stand and honouring a theory which still lacked five of its 
essential particles --- the $W^\pm$ and $Z^0$ bosons, the Higgs boson, the top quark and the tau neutrino. 
Their confidence was brilliantly reinforced by the 1983 discovery, at CERN's Super Proton Synchrotron 
(SpS), of the $W^\pm$ \cite{11, 12} and $Z^0$ \cite{13} bosons, which prompted another Nobel Prize, the 
very next year, to Carlo Rubbia and Simon van der Meer, who were the leaders of this experiment. This 
marked the beginning of the reign of the Standard Model (SM), a name which had been coined in 1975 by 
Treiman and Pais \cite{15}. That the SM would prove so durable was not, however, quite apparent at the 
time, and the literature was simply buzzing with alternative ideas.

The theoretical idea which attracted the most attention in the early 1980s was definitely that of grand 
unification, which was inspired by the success of the electroweak unification achieved by Glashow, Salam 
and Weinberg. Proposed by Georgi and Glashow \cite{16} in 1975, building on pioneering work by Pati and 
Salam (1974) \cite{17}, this sought to unify the strong interactions with the electroweak, within the 
framework of a single spontaneously-broken gauge theory. However, it was recognised that any such theory 
would immediately lead to fast decay of protons, unless the scale of the symmetry-breaking lies around at 
least $10^16$ GeV\footnote{The lower bound on the proton decay then was $3\times 10^{30}$~years, as 
determined in Ref. \cite{18}. It has since risen by about four orders of magnitude. For the latest 
results, see Ref. \cite{19}.}. This immediately brought to the fore a problem with the Higgs boson. The 
fact that the mass of an elementary scalar is not stable under quantum corrections, but tends to rise to 
the cutoff scale for the theory, had been recognised by Wilson as early as 1971 \cite{20}. In Grand 
Unified Theories (GUTs), this cutoff scale would then be around $10^{16}$~GeV --- clearly an intolerable 
situation for the Higgs particle whose mass was already known to be restricted by perturbation theory to 
lie within a TeV, and is now known, thanks to the LHC, to be around 125~GeV. This problem acquired the 
name of the 'gauge hierarchy problem' \cite{21}, or simply, the hierarchy problem.

It may be noted in passing that the hierarchy problem does not exist if the Standard Model holds all the 
way to the Planck scale, the energy scale around $10^{19}$ GeV at which the effects of spacetime 
curvature, i.e. gravity, become as strong as the gauge interactions which describe the electroweak and the 
strong interactions. For then there exists a well-known mechanism -- the mass renormalisation technique -- 
to absorb the large quantum corrections into the bare mass of the Higgs boson, leaving a small quantity 
which is measurable. However, if there is an intermediate scale, like the GUT scale, where there are 
physical fields with physical masses, then, even if the same huge quantum corrections are absorbed into 
the Higgs mass at one energy scale, they will reappear (somewhat diminished but still intolerably large) 
when the physical masses run to another scale. In the 1980s, when most high-energy physicists believed 
that a GUT of some form does exist, this was the obvious situation to be faced, with the Higgs boson mass 
getting dragged up to the GUT scale. Today, we are not quite so sure \cite{22}.

Two different solutions of the hierarchy problem had emerged by the end of the 1970s. One was a rather 
well-traversed idea, based on the behaviour of strong interactions, which bind the quarks into hadrons. 
These can be split into other hadrons, but cannot be broken back into quarks --- a phenomenon called quark 
confinement. The strong interaction (QCD) coupling, in fact, grows stronger as the energy scale falls and 
appears to diverge at a value around a quarter of the nucleon mass, the so-called QCD 
scale\footnote{Technically, we would call it a Landau pole in the running of the strong coupling constant, 
and it really indicates that the interaction has become too strong to be described correctly by 
perturbation theory.}. It is this nonperturbative interaction that binds quarks into hadrons, and hence 
the scale of hadron masses is set by the scale of this breakdown point \cite{23,24}. Now hadron fields are 
known to form Lorentz scalars, as well as vectors, tensors and fermions. Could the Higgs boson be just 
such a composite object, bound by a much stronger (maybe gauge, like QCD?) interaction, where the scale 
(or Landau pole) lies around the Higgs boson mass? This, in a nutshell, is the contention of a class of 
theories collectively dubbed technicolor theories \cite{25,26}. The beauty of such theories lies in the 
fact that the Standard Model breaks down at a scale around a TeV, or a few TeV, and the new gauge 
interaction comes into play. Quantum corrections to the Higgs boson mass would, then be completely 
manageable and there would be no hierarchy problem. The foundations of this theory had been laid down as 
early as 1973, by Jackiw and Johnson \cite{27}, and independently by Cornwall and Norton \cite{28}. 
However, it was the 1976 work of Weinberg \cite{29}, describing the gauge field theories necessary to 
drive technicolor, followed by the 1978 work of Susskind \cite{30}, applying this to the electroweak 
model, which made technicolor a viable means of theorising beyond the Standard Model. Obviously, such a 
model would make naive grand unification redundant, it being assumed that the electroweak and strong 
interactions are effective theories in the same way as Yukawa theory is an effective theory of 
QCD\footnote{In fact for hadronic physics, this idea has been developed into what is called chiral 
perturbation theory \cite{31,32,33}.}.
 
Compositeness is always an attractive idea, given that atoms, nuclei and hadrons were all found to be 
composites as soon as experiments began to probe inside them. The ideas of primitive technicolor were soon 
taken ahead by Harari \cite{34} and, independently, by Shupe \cite{35}, who developed the first so-called 
'preon' models, where all the Standard Model particles are taken to be composites, not just the Higgs 
boson. Preon models flourished in the 1980s, with one of the most dedicated exponents being Pati 
\cite{36}, whose 1974 paper with Salam \cite{17} contained one crucial sentence which inspired the whole 
genre of preon models.

By far the most popular solution of the hierarchy problem, however, was supersymmetry (SUSY). It is still 
a popular model, though somewhat the worse for wear after the second run of the LHC. The inspiration for 
this came from string theory, where the basic excitations are all bosonic, and, therefore, one needs 
supersymmetry to have fermions \cite{37,38,39}. It was already known from the 1975 work of Haag, 
Lopuszanski and Sohnius \cite{40, 41} that if we want to have a super unified theory, where gravity is 
united with the gauge interactions, then this will have to be supersymmetric. However, this might never 
have become relevant in the LHC context, had it not been for the discovery that the quadratic divergences 
in the Higgs boson self-energy cancel out in SUSY theories, leaving behind only logarithmic divergences, 
which are easy to absorb into the measured mass. This 'miracle' was discovered in the simplest SUSY model 
by Wess and Zumino in 1974 \cite{42, 43} and extended to a SUSY version of QED. Extension to a nonAbelian 
gauge theory was achieved a year later by Ferrara and Zumino \cite{44, 45}. In 1976, Pierre Fayet wrote 
down a SUSY version of the Standard Model \cite{46}, but it was not till the work of Dimopoulos and Georgi 
(1981) that it was established that the SUSY must be explicitly broken to be phenomenologically viable 
\cite{47}. Soon after, several authors established the cancellation of quadratic divergences in this 
so-called Minimal Supersymmetric SM, or MSSM. Among them were two Indians, viz. Kaul (then at the IISc, 
Bengaluru) and Majumdar (then at the TIFR, Mumbai), who published a series of papers \cite{48,49,50}, 
discussing the cancellation of quadratic divergences in anomaly-free SUSY gauge theories, including the 
MSSM. Though not directly connected with the LHC, these may be regarded as one of the first important 
contributions from India to physics which was -- and still is -- central to the analysis of LHC data.

In the 1980s, therefore, the efforts of theoretical particle physicists centred around four major themes.
  
\vspace*{-0.2in}
\begin{itemize}

\item One was the Standard Model itself and its hitherto undiscovered components --- the top quark, the 
Higgs boson and the tau neutrino. In this genre, we may include bottom-up extensions of the Standard 
Model, such as a fourth generation of fermions \cite{51,52}, exotic fermions \cite{53,54,55,56,57,58}, 
two-Higgs doublet models \cite{59,60,61}, extra gauge bosons ($W'$ and $Z'$) \cite{62, 63}, and left-right 
symmetry \cite{64,65,66,67,68,69}. The study of neutrinos and neutrino oscillations may also be included 
in this category \cite{70,71,72,73}.

\item The next was GUT model building \cite{74, 75}, which sought to understand various features of the SM 
masses and mixings as consequences of higher gauge symmetries and their spontaneous-breaking patterns. 
This received a major boost from string theories, which brought into the limelight large symmetry groups 
such as $E_8 \times E_8$ \cite{76}, $E_6$ \cite{77, 78} and $SO(10)$ \cite{79, 80}, as well as large 
representations, such as the 126-plet of Higgs bosons in $SO(10)$ models. These may otherwise have lain 
beyond the wildest dreams of particle phenomenologists.

\item The study of compositeness, especially technicolor, was pursued by some small but prolific groups of 
particle physicists, the most well-known group being centred around Harvard University and led by Howard 
Georgi, Ken Lane, Bill Bardeen, and others. Technicolor models predicted the discovery of technipions and 
techni-rho particles at colliders, which would look like heavy scalars or vector bosons \cite{81, 82}. 
Preon models were somewhat peripheral to the mainstream literature, as they were thought of then -- as now 
-- as premature and somewhat speculative \cite{83}. \item The study of supersymmetry, especially the MSSM, 
and its most restrictive version, then known as the minimal SuperGravity Model (mSUGRA) \cite{84} and 
later as the constrained MSSM, or cMSSM, grew enormously in this period, and may have reached its climax 
in the 1990s. The most popular clich\'e in this context was to describe the MSSM as the 'standard model' 
of beyond-SM physics. It was during this period that the well-known collider signals \cite{85,86,87}, like 
trileptons, missing energy and like-sign dileptons, were predicted and studied. \end{itemize} 
\vspace*{-0.2in} It was widely expected in the 1980s that just as the SM had been vindicated at the SpS, 
some signal from these new theories would be found at the next generation of machines, viz. the LEP at 
CERN and the Tevatron at FNAL. After that would come the third generation of colliders, viz. the LHC at 
CERN and the SSC in Texas. It was definitely time to start thinking about the physics of these futuristic 
machines \cite{88}.

In the following discussion, we will describe Indian contributions to these themes under different 
subheadings, together with brief notes on the general developments in those areas. Of course all of these 
were being pursued at the same time -- with often the same person pursuing more than a single theme -- so 
that such a description will not be strictly linear or follow a rigid timeline. Moreover, over and above 
the above list, new ideas surfaced towards the turn of the century, such as brane worlds, unparticles and 
little Higgs models. The lack of evidence for any of these has created a hostile environment when 
researchers take shelter within the bunker of effective field theories and simplified bottom-up approaches 
to LHC physics. At the same time, there have been momentous developments in neutrino physics, and in 
cosmology and astroparticle physics, which will not be described in any detail in this article but will 
still find some mention in passing, for they have influenced LHC studies as, indeed, all of high energy 
physics. At the cost of repeating a well-worn clich\'e, it is still worth stating that scientific research 
is not conducted in a vacuum. Nevertheless -- and this may be treated as a disclaimer -- it has not been 
possible to include all Indian efforts in the list of citations in this article. Nor has that been the 
purpose of the albeit long list of references, which is for the purpose of illustration and guidance, and 
is not intended to form a compendium. Furthermore, this article does not describe the interesting, even 
exciting, results being found at the LHCb and ALICE experiments --- partly for lack of space, but also 
because the author lacks expertise to comment on these in any insightful manner.

One final remark before we embark on our voyage of discovery. In this article, the work 'Indian' has been 
used in a fairly general sense to include all high-energy physicists of Indian origin who are working or 
have worked in India in any capacity, whether as graduate student, postdoc or faculty member. The overseas 
work of such individuals has also been considered. The work of Indians and scientists of Indian origin who 
have settled abroad have, by and large, been omitted, with some notable exceptions.

\section{Taming the top}\label{sec2}

It is now generally accepted that the prediction of a third generation of fermions was made as early as 
1973 by Kobayashi and Maskawa \cite{89}. Their work, even though it eventually won the authors a Nobel 
Prize, did not make much of an impact at the time it was proposed. Thus, when the tau lepton ($\tau$) was 
discovered at SLAC in 1975 \cite{90}, it came as something of a surprise. However, within 2 years, in 
1977, the bottom quark ($b$) was also discovered at Fermilab \cite{91}. There was, thus, an incomplete 
generation, where it was assumed that there was a leptonic and a quark doublet of $SU(2)_L$ with missing 
partners --- dubbed, for obvious reasons, the tau neutrino ($\nu_\tau$) and the top quark ($t$), 
respectively. The standard argument for the existence of these then-undiscovered fermions arose from the 
requirement of anomaly cancellation in the Standard Model \cite{92, 93}. More specifically, the 
requirement is that $\sum_f N_f Y_f = 0$ where $Y_f$ is the hypercharge of the fermion and $N_f$ is the 
colour multiplicity (1 for leptons, 3 for quarks), and the sum runs over one generation. The cancellation 
must happen over each generation separately. Taking the Standard Model values for the third generation, 
i.e. $Y(\nu_\tau) = -1$, $Y(\tau_L) = -1$, $Y(\tau_R) = -2$, $Y_{tL} = 1/3$, $Y_{bL} = 1/3$, $Y_{tR} = 
4/3$ and $Y_{bR} = -2/3$, and appropriate colour factors, it is easy to verify that the required 
cancellation does occur. Obviously, it will not happen if one or more of the fields are missing. This was 
the most compelling reason to believe that the missing top quark and the missing tau neutrino must exist. 
However, there did exist speculations which assigned different quantum numbers to these fermions, e.g. 
making the $b_L$ a singlet of $SU(2)_L$. Most of these 'topless models' \cite{94} were ruled out by data 
from the early runs of the LEP Collider at CERN, which started running in 1989. However, very similar 
ideas are now being reworked in the context of an extra $Z'$ in models beyond the SM.

The fact that a high-energy hadron collider would be an ideal place to look for a top quark, because it 
can be produced in strong interactions, was already recognised by 1983, when the first Indian paper on the 
topic was published by the trio of Roy, Godbole and Pakvasa \cite{95}. At the time, it was already known 
that the mass of the top quark must be more than about 20 GeV, since the PETRA -- an $e^+e^-$ collider at 
DESY, operating at a centre-of-mass energy between 27 and 46 GeV -- had failed to see any signal for 
top-quark pair production. In their pioneering work, Roy {\it et al.} pointed out that the semileptonic 
decay of the top quark, viz. $t \to b + W^+ \to b + (\ell^+ \nu_\ell)$ will lead to a final state with a 
hard transverse lepton $\ell = e, \mu, \tau$ which is isolated from the jet which will originate from the 
b quark in the decay. In their studies, isolation meant a minimum angular distance $\Delta R > 0.4$, where 
$\Delta R^2 = \Delta \eta^2 + \Delta \varphi^2$, in terms of the pseudorapidity difference $\Delta \eta = 
\eta_b - \eta_\ell$ and the azimuthal angle $\Delta\varphi = \varphi_b - \varphi_\ell$ between the lepton 
$\ell$ and the thrust axis of the $b$-jet\footnote{These quantities remain invariant under longitudinal 
boosts and make a better measure than the opening angle. Nowadays, other invariant measures of distance, 
such as $k_T$ are more popular.}. The point made by Roy {\it et al.} -- which was also independently 
arrived at by Barger {\it et al.} \cite{96} around the same time -- is that in the similar semileptonic 
decays of the bottom or charm quarks, which are much lighter, the decay products will be much more 
strongly boosted along the direction of the parent particle, leading to a non-isolated lepton, i.e. which 
will appear as part of the jet. Thus, the heavy mass of the top quark, which renders its production 
difficult (well-nigh impossible at the energies available in the 1980s), also leads to a clear signal, 
once it is produced.

The isolated lepton signal dominated the top-quark searches at Fermilab's Tevatron machine, which ran at 
1.8 (2.0) TeV from 1987 till 2011. It was also the channel through which the top quark was finally 
discovered by the CDF and D0 Collaborations at the Tevatron in 1995. Meanwhile, in 1984, Roy alone 
\cite{97}, and also in collaboration with Lindfors \cite{98}, had shown that electroweak production of top 
quarks at a hadron collider is suppressed by an order of magnitude. A detailed analysis of some candidate 
events seen at the UA1 detector at CERN's Super Proton Sychrotron (SpS) was published by Roy with Godbole 
\cite{99}, in which they concluded that both semileptonic and fully hadronic (three-jet) decay modes of a 
top quark of mass around 35 GeV could explain the candidate events. However, these events were unlikely to 
have originated from top quarks, since the SpS energy was only about 500 GeV, where we know with hindsight 
that the probability of producing a pair of 172 GeV top quarks is negligibly small \cite{100}. In fact, 
the first results on $B^0$-$\bar{B}^0$ mixing -- which came in 1987 from the ARGUS Collaboration at the 
DORIS II ring at DESY, Hamburg -- immediately showed that the top quark mass must lie above 60 GeV in the 
Standard Model \cite{101}. Five years later, the early runs from the Tevatron at Fermilab had pushed the 
lower bound to 91 GeV \cite{102}. By 1994, the lower bound rose to 131 GeV \cite{103}.

At around the same time, the LEP Collider began to run at CERN \cite{2}. It had been hoped that it would 
lead to the rapid discovery of the Higgs boson, but the elementary scalar proved frustratingly elusive. 
However, LEP did provide accurate measurements of the masses of the $Z$ and the $W$ bosons, which are 
affected at the loop level by both the Higgs boson and the top quark. Within 2-3 years of running, global 
fits to the LEP data were able to push the top-quark mass limit to somewhere between 150 and 190~GeV 
\cite{104}. These, of course, assumed that there are no particles other than the Standard Model ones. When 
the top quark was finally discovered at the Tevatron -- by the CDF Collaboration \cite{105}, followed some 
months later by the D0 Collaboration \cite{106} -- its mass came out as around 175 GeV, i.e. right in the 
middle of the expected range.

The top quark is definitely an odd fish among the fermions in the Standard Model. As it considerably 
heavier than its closest rival, the $b$ quark, it is the only parton which decays before it can hadronise. 
Another consequence of this large mass is that in the Standard Model the Yukawa coupling of a top-quark 
pair with the Higgs boson come out as almost exactly unity --- by far the largest coupling constant in the 
Standard Model at the electroweak scale. A third consequence is the large separation between the measured 
mass around 172~GeV and its MS mass, which is around 165~GeV. We may recall that like the other quarks we 
cannot define a pole mass for the $t$ quark, since quark states are never asymptotically free. The heavy 
mass of the top quark also makes it the principal driver of loop corrections in a variety of processes, 
including $B^0$-$\bar{B}^0$ mixing, the Higgs boson self-energy, and the RG evolution of the quartic 
scalar coupling $\lambda$ in the Standard Model. In models beyond the Standard Model, such as the 
constrained minimal Supersymmetric Standard Model, it is such top-quark-mediated loops which provide an 
elegant mechanism to drive electroweak symmetry-breaking. As a result, the top quark and its properties 
have been a matter of great interest to high energy physicists every since its discovery.

One reason why it is imperative to know the mass of the top quark precisely is mentioned above, which is 
its pivotal role in determining if the $\lambda$ parameter remains positive and the electroweak vacuum 
remains stable all the way to the Planck scale \cite{107}. For, if it does not, then there would be a 
high-energy scale where all particles would go down to indefinitely large negative energies and emit huge 
amounts of radiation, leading to a 'fiery death' of the Universe. Conversely, since the Universe began in 
such a 'fiery' situation, well above the scale of instability, it could never have cooled down close to 
absolute zero --- which it has. Therefore, one would need to postulate the existence of new physics beyond 
the Standard Model in order to provide the required vacuum stability. With the present level of accuracy 
of top quark mass measurements, however, we cannot yet make a definitive statement on this issue 
\cite{107}.

The mainstream in top-quark physics at the LHC has been the measurements of the production cross-section 
and the couplings of the top quark to the gauge bosons, as well as the Higgs boson. Theoretical studies of 
these have centred around calculating higher order QCD corrections to the production cross-section of 
$t\bar{t}$ \cite{108} as well as $t\bar{t}X$ where $X = Z,H$ \cite{109, 110}. The Indian community has, 
however, focussed on the top quark as a portal to BSM physics. The pioneer in this has been Rindani, who 
has worked on polarisation effects, mostly in the context of $e^+e^-$ colliders from 1995 onward 
\cite{111}. In 2010, he also teamed up with Godbole {\it et al.} \cite{112} to discuss the LHC signals for 
nonstandard couplings which would generate a nonzero top-quark polarisation, showing that asymmetries in 
the azimuthal angle distribution of leptons arising from top semileptonic decays can be a powerful probe 
of such effects. Top polarisation studies were then carried forward by Godbole, with her collaborators 
\cite{113, 114}, mainly analysing the $t\bar{t}H$ coupling and developing angular variables where BSM 
effects would show.

Detailed studies of nonstandard top-quark couplings in one-loop effects have been carried out by Choudhury 
and Saha, who explored chromomagnetic and chromoelectric dipole interactions of the top quark in a 
model-independent framework \cite{115} and then the possible effects of anomalous top-quark couplings in 
Higgs boson production at the LHC \cite{116}. On a slightly different note, Bardhan {\it et al.}, 
discussed flavour-changing decays of the top quark at one loop, including the process $t \to c + Z^0$, 
setting up a general framework in which BSM predictions could be compared with the data \cite{117}.
 
The top quark has also featured in other kinds of BSM studies, where it is generally a decay product of 
some heavier nonstandard particle. Indian contributions in these directions have been described in the 
following sections. However, this section would not be complete without a mention of the new paradigm for 
top-quark identification at the LHC, namely jet substructure. It was pointed out by Kaplan {\it et al.} in 
2008 \cite{118} that hadronically decaying top quarks could be tagged by a consideration of the jet 
substructure, since the final-state jets at the LHC energies would be likely to be highly boosted, 
compared to earlier machines. This has since become a standard technique. However, another obvious decay 
channel in this context is the semileptonic decays, originally touted as the 'smoking gun' signals for top 
quarks, but which remained unexplored in the context of boosted top decays. This lacuna was filled in by 
Chatterjee {\it et al.} \cite{119}, who, in 2019, developed a technique to tag 'electron-rich' jets as 
originating from a top-quark decay --- or otherwise.

\section{Hunting for Higgses}\label{sec3}

The Higgs boson has been described as the keystone of the Standard Model, and its discovery \cite{9, 10} 
in 2012 was one of the greatest triumphs of modern science, both theoretical and experimental. 
Nevertheless, this success came only at the end of a protracted search lasting several decades. Being the 
only elementary scalar in the SM, the Higgs boson has been the subject of intense speculation, from its 
mass, spin and magnetic moment, to its couplings with all the other SM fields, as well as its vacuum 
structure. Something of this massive level of questioning can be gauged by a perusal of the 
(somewhat-dated) monograph The Higgs Hunter's Guide \cite{59,60,61}, which provides a comprehensive 
introduction to the subject as it was in the 1990s.

While the role of a Higgs doublet in engineering spontaneous symmetry-breaking in the SM gauge sector had 
been worked out in the 1960s, it was the discovery of the $W$ and $Z$ bosons which made the presence of a 
Higgs boson imperative, for without it the scattering amplitude for, say, $e^+e^- \to WW$ becomes 
non-unitary in the high-energy limit. In fact, purely from unitarity considerations one could infer that 
the Higgs boson would have a mass within about 1 TeV. At that stage the unitarity problem could have been 
explained away by invoking a strongly-interacting electroweak sector \cite{120}, but such hopes faded away 
when such models became severely constrained by the data from LEP in the 1990s. Once the top quark mass 
became known in 1994, the 1998 global fit of all electroweak data by Erler and Langacker \cite{121} also 
brought the Higgs boson mass down to below 225 GeV. The direct search data at LEP had already established 
114 GeV as a lower bound. Thus, by the turn of the century, a discovery within this narrow window by the 
LHC was more-or-less guaranteed, so far as the SM was concerned. As a matter of fact, this goal was 
achieved rather early in the LHC run, with the 2012 discovery

As with the top quark, Indian contributions to the physics of Higgs bosons in the LHC context has mostly 
been in the area of BSM models. A notable exception, however, has been the crucial work of Ravindran, who, 
with his collaborators, has been steadily working out QCD corrections in the SM to the dominant $g + g \to 
H^0$ partonic process for Higgs production at the LHC --- with increasing accuracy over the years. These 
corrections were at NLO in 2002 \cite{122}, at NNLO in 2003 \cite{123}, at full two loops in 2005 
\cite{124}, with threshold corrections beyond NNLO in 2007 \cite{125}, at NNNLO in 2014 \cite{126}, and at 
NNLO + NNLL in 2018 \cite{127} and 2021 \cite{128, 129}. Further efforts include similar series of 
calculations of the $b + b \to H^0$ process to three loops and beyond \cite{130,131,132,133,134} and 
calculations of the $g + g \to H^0 + H^0$ process \cite{135,136,137}, the latter being vital for the 
direct measurement of the quartic Higgs couplings $\lambda$ in future runs of the LHC. It may be noted 
that not only do these extremely sophisticated calculations make the SM prediction more accurate, but even 
in variables (such as Higgs signal strengths) when the cross-section eventually cancels out, these make 
the result almost independent of the scale of renormalisation.
      
In the field of BSM Higgs studies, there have been a few studies of exotics such as singlet \cite{138}, 
triplet \cite{139} and composite Higgses \cite{140}, but most of the Indian work has centred around 
two-Higgs doublet models (2HDM), especially with regard to the additional charged Higgses and the 
pseudoscalar Higgs predicted in these models. Moreover, since SUSY models are 2HDMs with a severely 
constrained parameter space, a fair portion of the 2HDM analyses have been carried out in a SUSY 
framework. An interesting early study of charged Higgs bosons by Biswarup Mukhopadhyaya and collaborators 
identified the possibility of observing radiation amplitude zeroes in the production and decay of a 
charged Higgs boson \cite{141}. Though written in the context of the defunct SSC, the result is equally 
applicable to some processes at the LHC \cite{142}.

Nevertheless, the pioneer in detailed collider studies of nonstandard Higgs bosons, as in the case of the 
top quark, was undoubtedly Roy, who initially considered a charged Higgs boson $H^\pm$ decaying into a top 
quark, both at the partonic level \cite{143} and with QCD corrections \cite{144}. Then it was the 
pioneering paper of Bullock {\it et al.} \cite{145}, in which they pointed out that the polarisation of 
$\tau$ leptons arising in $H^\pm$ decays is different from that of those arising from $W^\pm$ decays 
(their principal SM background at hadron colliders including the LHC) which inspired a series of articles 
from Roy and his collaborators, in which this idea was explored with ever-increasing degrees of 
sophistication \cite{146,147,148,149,150,151,152,153}. On a variant note, Roy, with his collaborators, 
also studied a possible multi-$b$ quark signal for the $H^\pm$ \cite{154,155,156} and $\tau$ polarisation 
for the pseudoscalar in 2HDMs \cite{157}, as well as the rarely studied process $H^\pm \to W^\pm h^0$ 
\cite{158}. Other studies of charged Higgs bosons include a study by AseshKrishna Datta and collaborators 
on H± produced in cascade decays of SUSY particles \cite{159}. Novel signatures for $H^\pm$ have also been 
explored in the context of extra Yukawa couplings \cite{160} and a chromophobic $H^\pm$ in a 
$Z_2$-extension of the SM \cite{161}. Similar studies of neutral scalar and pseudoscalar Higgs bosons in 
the context of stop decays in SUSY models have been explored in 2000 by Sridhar \cite{162}, in 2015 by 
Mukhopadhyaya \cite{163} and by Ghosh \cite{164}, each with their respective collaborators.

With the 2012 discovery of the obviously-neutral scalar of mass 125~GeV at the LHC came immediate 
questions about the spin and parity of this new particle. In a 2011 study \cite{165}, Mukhopadhyaya and 
collaborators had already showed how final states in Higgs production via vector boson fusion could 
provide the clue to this, as well as indicate the presence of anomalous HWW vertices. Post-Higgs discovery 
analyses on similar lines were carried out by Godbole with her collaborators in the context of the LHC 
\cite{166, 167} and a possible LHeC \cite{168}. In all these studies, the azimuthal angle distribution of 
final states is the key to identification of anomalous couplings.
           
In the days preceding the Higgs boson discovery, the absence of Higgs signals engendered fears that the 
Higgs may actually have been produced and remained undetected because it dominantly decays 'invisibly'. An 
early idea was a Higgs boson decaying into heavy neutrinos \cite{169, 170}, but this possibility receded 
with the LEP data. LHC signals for an invisibly decaying Higgs were also explored fairly early by Roy and 
collaborators \cite{149, 171}. Post-discovery, the possibility of a subdominant but measurable 'invisible' 
branching fraction of the observed Higgs was investigated by Ghosh {\it et al.} in 2013 \cite{172}. 
Invisible decay modes of the Higgs boson have become highly relevant today in the context of dark matter 
searches, though all we have, till now, is a fairly stringent upper bound \cite{173}. As a matter of fact, 
after the Higgs boson, dark matter has become the new 'Holy Grail' of high-energy physics, and the Higgs 
boson is often considered as a portal to this unknown world. In this context, Datta {\it et al.} made a 
detailed study of multi-lepton signatures \cite{183} of the 'inert doublet model' where a decoupled scalar 
doublet plays the role of dark matter \cite{174}. Several other studies on similar lines followed 
\cite{174,175,176,177,178,179,180,181,182}.
   
The discovery of the Higgs boson has also created renewed interest in the electroweak vacuum and eventual 
fate of the Universe. Ever since Coleman and Weinberg \cite{184} pointed out the scale-dependence of the 
shape of the electroweak potential, the fate of the electroweak vacuum has been a matter of intense 
speculation. However, no definite statements could be made since the mass of the Higgs boson was not known 
at any scale. Once this became known in 2012, a clutch of articles \cite{185,186,187,188} appeared, 
pointing out the possibility that the electroweak vacuum might destabilise at a scale somewhere between 
$10^{11}$~GeV and the Planck scale.

Vacuum stability was soon recognised to be a powerful tool to probe BSM physics, and several Indian groups 
took this up. Anindya Datta and Sreerup Raychaudhuri studied universal extra dimensions \cite{189, 190}, 
while neutrino mass models were studied by Chakrabortty {\it et al.}, who considered Majorana neutrinos 
\cite{191}, Khan {\it et al.}, who considered a minimal seesaw model with a scalar singlet \cite{192} and 
Bhupal Dev {\it et al.}, who considered a Type-II seesaw \cite{193}. The SUSY Higgs vacuum was considered 
by Chowdhury {\it et al.} \cite{194, 195}, a similar non-SUSY model was discussed by Chakrabarty {\it et 
al.} \cite{196}, and a singlet scalar dark matter candidate was considered by Khan and Rakshit \cite{176}. 
Other work of similar nature followed, for vacuum stability has now become a standard constraint to check 
for any new physics model involving the Higgs sector.

\section{Seeking supersymmetry}\label{sec4}

The rise and fall of supersymmetry (SUSY) must, when histories of the current period are written, become 
the dominant theme of the 3 decades from 1990–2020. For no theory of elementary particles -- not even the 
SM -- has come packed with so many phenomenological goodies as SUSY. Let us quickly list the main ones.

On the aesthetic side, SUSY not only does away with the ugly distinction between bosons and fermions which 
is a legacy of quantum mechanics, but also provides the only way to unify gravity with the other 
interactions in a gauge theoretic framework \cite{40, 41}. On the practical side, we have already 
discussed how it eases the Hierarchy Problem in GUTs \cite{48,49,50}. Great excitement arose in 1991 when 
Amaldi {\it et al.} \cite{197} showed that when measurements of the three SM coupling constants at the 
$Z$-pole were combined with low-energy measurements, the running of the couplings indicated that there was 
no scale where all three would become identical, as predicted in a GUTs. However, when a SUSY $SU(5)$ GUT 
model was taken, with the threshold for sparticles around 1~TeV, proper unification could be achieved at a 
scale around $10^{16}$~GeV. Since then, the bounds on proton decay have pushed up the unification scale 
roughly an order of magnitude higher \cite{198}, but this can be accommodated by a change in the GUT group 
\cite{199, 200}. SUSY also permits the running of the mass parameters in the Higgs sector. A beautiful 
feature of this is that one of these mass-squared parameters can go negative, breaking the electroweak 
symmetry at precisely the right scale. To achieve this, one requires a Higgs boson mass less than about 
135~GeV --- a robust prediction of SUSY models \cite{201,202,203,204,205,206}. Indeed the Higgs boson does 
have a mass of 125~GeV, which is a tantalising, but in no way compelling, agreement with SUSY. Last, but 
not least, SUSY models can incorporate an elegant $Z_2$ symmetry called $R$-parity which makes the 
lightest supersymmetric particle (LSP) both stable and invisible --- the perfect WIMP candidate for dark 
matter \cite{207}.
 
Given the runaway popularity of SUSY as the canonical BSM theory, it has come as a huge shock and surprise 
to the high-energy physics community that even a decade after the 2012 Higgs triumph, the LHC has not 
found a shred of evidence for supersymmetry \cite{208, 209}, or indeed, of anything which can definitely 
be described as BSM \cite{210}. This has led to a reaction of sorts, with supersymmetry sometimes being 
castigated as 'a nightmare scenario' \cite{211} with some even calling for a total rethink of the 
traditional synergy between mathematical theory and experiment \cite{212}. However, such pessimism is 
premature, as has been pointed out by many experts, including review articles in the SUSY context by Datta 
\cite{213} and by Mukhopadhyaya \cite{214}. Nevertheless, the focus of the high-energy community has 
definitely drifted away from SUSY towards areas such as dark matter, effective field theories and minimal 
models, which are touched upon in Sect.~6.

Before proceeding further, it is necessary to mention the nomenclature of different low-energy SUSY 
scenarios. Since supersymmetry must be broken, one has to introduce a slew of 'soft' SUSY-breaking terms, 
which will give masses to the superpartners, making them heavier than their SM counterparts. This leads to 
an explosion in the number of free parameters in the theory. The MSMM has 37 $CP$-conserving free 
parameters—and another 67 if $CP$ is violated. As a result, a Higgs boson below about 135~GeV 
\cite{201,202,203,204,205,206} is perhaps the only robust prediction of SUSY models. Theorists, therefore, 
have tried to model the SUSY breaking through different mechanisms, each idea individually leading to a 
drastic reduction in free parameters --- at the cost of a proliferation of models. The earliest of these 
was a minimal supergravity, or mSUGRA framework, which assumed a GUT and a 'hidden sector' where SUSY is 
broken, presumably spontaneously \cite{84}. Combining this with a mechanism for electroweak 
symmetry-breaking produced the so-called 'constrained' MSSM, or cMSSM \cite{215,216,217}. Some of the GUT 
constraints were relaxed to produce the so-called 'non-universal' models \cite{218}. Alternative 
mechanisms for SUSY-breaking led to the 'gauge-mediated' SUSY breaking (GMSB) \cite{219} and 
'anomaly-mediated' SUSY breaking (AMSB) models \cite{220, 221}. This was followed by the idea of 'split 
SUSY', with an unabashedly 'unnatural' mass spectrum \cite{222,223,224,225}. Null results at the LHC have 
prompted the exploration of various 'phenomenological' MSSM, or pMSSM versions, with subsets of the 37 
parameters being considered independent -- typically 10, 15, or 21 of them -- while the rest are either 
fixed or set to zero \cite{226,227,228}.

As may be expected, a theory, or, rather, class of models, with so much to offer rapidly became the 
mainstream for BSM physics in the 1990s and continued, as we see, till fairly recently. As with the rest 
of the world, contributions proposing detailed or novel signatures for SUSY in all its avatars form the 
bulk of high-energy phenomenology contributions from India. The scope of this article does not permit a 
comprehensive description of this. Hence, in the discussion that follows, a general portrayal will be 
attempted, highlighting some specific works and perforce omitting many, many others.
     
The pioneering Indian SUSY studies in the context of hadron colliders were made, as in other areas, by Roy 
and his collaborators, starting from the 1980s. In particular, Roy and Reya wrote a series of papers 
\cite{229,230,231,232,233} highlighting the fact that the Majorana nature of gluinos would result, when 
their cascade decays into final states with leptons are considered, in like-sign dilepton signals. Roy 
continued these studies with Drees {\it et al.} \cite{234,235,236,237,238,239,240,241,242} till the advent 
of the LHC. On another note, he carried his experience of using $\tau$ polarisation in H± decays to SUSY 
signals, especially with staus \cite{243,244,245,246}. These and similar ideas were carried ahead by 
Godbole with B\'elanger and her other collaborators \cite{247,248,249} in the changing scenario as the LHC 
(null) results started coming in. Datta and his collaborators were among the earliest to point out that 
'invisible' decays of other SUSY particles, such as the next-to-LSP or NLSP, could enhance the 
missing-energy signals from the LSP. With the increasing automatisation of cascade decay calculations, 
such pinpointed studies have merged into the general SUSY searches for invisible decays or, equivalently, 
SUSY dark matter searches \cite{250,251,252,253,254,255,256,257,258,259,260}. Datta, with his 
collaborators also carried out a series of studies of the 'non-universal' varieties of SUSY, under 
different ans\"atze for the soft SUSY breaking parameters \cite{261,262,263,264,265,266,267,268,269}, 
culminating, as may be expected, in the pMSSM \cite{270}, as well as a series of studies of stability of 
the electroweak potential within SUSY models and corresponding constraints on the SUSY mass spectrum 
\cite{271,272,273}.

Many different aspects of SUSY models were studied over the years in the school set up by Mukhopadhyaya 
with his students and collaborators. An exhaustive list would, perhaps, require a separate article, but 
some of the LHC-related work is highlighted here to indicate the range and variety of their oeuvre. One of 
the earliest of these studies was the consideration of vector boson fusion to produce charginos and 
neutralinos, as well as sleptons at the LHC in the MSSM as well as in AMSB models 
\cite{274,275,276,277,278,279}. Several studies of split SUSY followed \cite{280,281,282,283,284}. The 
impact of neutrino masses and mixings on SUSY models and corresponding collider signals was explored over 
the years in a series of papers \cite{161,285,286,287,288,289,290,291,292,293,294,295,296,297,298,299}. 
Another topic of interest was 'non-universal' SUSY models \cite{292, 300,301,302,303,304,305}, in addition 
to a collection of more purely phenomenological investigations 
\cite{306,307,308,309,310,311,312,313,314,315, 316,317,318}, as the focus of the subject shifted away from 
UV-complete model building to the study of low-energy mass spectra. This inevitably has led to multiple 
studies of dark matter in the SUSY context \cite{163,288,297,299,319,320,321,322,323,324,325,326,327}. In 
addition, many other ramifications of SUSY models in the LHC context have also been discussed over the 
years by Choudhury \cite{328,329,330,331,332}, Sridhar \cite{162,333,334,335}, Raychaudhuri 
\cite{336,337,338,339,340,341,342} and Guchait 
\cite{343,344,345,346,347,348,349,350,351,352,353,354,355,356}, followed by Ghosh 
\cite{357,358,359,360,361,362,363,364,365,366}, Datta \cite{367,368} and Datta 
\cite{369,370,371,372,373,374,375,376,377,378,379}, with their respective students and collaborators. 
Younger entrants to this field include Rai \cite{380,381,382,383,384,384,385,386,387}, Konar 
\cite{388,389,390} and Bhattacherjee \cite{391,392,393,394,395,396,397,398,399,400,401,402,403,404,405}. 
It would take a full volume even to list the abstracts of these papers, indicating the enormous effort 
which has gone into the study of supersymmetry. Further, noting that the Indian SUSY oeuvre is merely a 
microcosm of the worldwide efforts (of which it is a small part) helps to realise that the experimental 
non-observation of SUSY has been little short of a d\'ebacle.

The Indian effort has not failed to touch two alternative viewpoints of supersymmetry, viz. the GMSB model 
\cite{357, 406,407,408,409,410,411,412,413,414,415} and the $CP$-violating version of SUSY \cite{308, 
416,417,418,419,420,421,422,423,424,425,426,427,428,429}, though it must be said that no studies have been 
undertaken in a coordinated or systematic fashion. On the other hand, the off-mainstream version of SUSY 
which seems to have had particular fascination for the Indian community is the version where R-parity is 
violated, i.e. there is no stable LSP, and hence no SUSY candidate for dark matter. Given that this 
removes one of the most attractive features of the theory, it is natural to ask why this less rewarding 
version has become so popular. The answer may be sought in the SM itself, where baryon number ($B$) and 
lepton number ($L$) are conserved, but purely accidentally, since they do not arise from the assumed 
symmetries, but just happen to be a symmetry of the Lagrangian when the other symmetries and 
renormalisability are imposed. The moment one tries to go beyond the SM, one encounters the problem that 
these may be violated. In SUSY, the second Higgs doublet superfield (required by holomorphicity of the 
superpotential) is forced (by the absence of tree-level flavour-changing neutral currents) to have the 
same gauge quantum numbers as the lepton doublet superfield, which opens up the possibility of mixing 
between these representations with different lepton numbers, i.e. violation of lepton number. To prevent 
this, which would lead to fast proton decay and rule out all SUSY models, $R$-parity was introduced 
\cite{430} as a compound of these symmetries, imposed on SUSY models. However, it was then pointed out by 
Weinberg and others, including Auluck and Mohapatra \cite{431,432,433} that proton decay requires 
violation of both baryon and lepton number and hence any one of the two can be violated while conserving 
the other. This opened up the possibility for $R$-parity violation. In such models, the LSP is no longer 
stable, but decays into multi-lepton final states if $L$ is violated, or, into multiple jets if $B$ is 
violated. Either way, one could predict spectacular signals at a machine like the LHC 
\cite{434,435,436,437,438,439,440,441}, and this is presumably where the Indian interest originated.

The Indian HEP community was introduced to the possibilities lying in R-parity violation by a talk given 
at the First Workshop on High Energy Physics Phenomenology (WHEPP-1) in Jan 1989 by Arnowitt \cite{442}, 
which was shortly followed by the publication of a magisterial analysis of the phenomenological 
implications by Barger {\it et al.} \cite{443}. As elsewhere, Indian contributions to this field 
originated with D.P. Roy, who studied the collider signals \cite{444,445,446,447}, and by P. Roy, who 
considered indirect signals \cite{448,449,450,451,452} for $R$-parity violation. The mantle was then taken 
up by others. The different aspects explored included novel signals at hadron colliders 
\cite{282,453,454,455,456,457,458,459,460,481,462,463,464,465,466,467,468,469,470,471, 
472,473,474,475,476,477,478,479,480,481,482}, collider bounds on $R$-parity-violating couplings 
\cite{484,485,486,487,488,489,490,491,492,493,494,495,496,497,498,499} and $R$-parity violation as an 
explanation for some (mostly transient) 'anomalies' in collider data \cite{500,501,502,503,504,505}. 
Several studies of bilinear $R$-parity violation were also made \cite{285,506,507,508,509,510,511,512}. 
Interestingly, even with the decline of interest in SUSY, $R$-parity-violating sparticles continue to be 
studied independently as the eponymous dileptons, diquarks and leptoquarks 
\cite{513,514,515,516,517,518,519,520,521,522,523,524,525,526,527,528,529,530}.

\section{Exploring extra dimensions}\label{sec5}

It has been related above that in the 1990s, it became quite clear that technicolor models could not be 
made to work without adding multiple ad hoc features, and, therefore, SUSY became the only plausible 
solution of the hierarchy problem in the SM. It was at this point (1998) that a new theoretical idea 
erupted on the high energy physics scene like a volcano which had long lain dormant. This was the idea of 
extra dimensions of spacetime.

By itself, the idea of extra dimensions was not new. An extra dimension of space had been dreamt of by 
philosophers \cite{531,532}, but it only became serious physics with the invention of Minkowski space and 
its subsequent application by Nordstr\"om {\it et al.} in separate attempts to unify Maxwellian 
electrodynamics with general relativity \cite{533,534,535,536,537}. These attempts died a natural death 
when they failed to correctly predict the $e/m$ ratio of known particles. However, extra dimensions were 
revived by the advent of superstring theories in the 1970s \cite{37, 538,539,540}. At this point, they did 
not seem to have much connection with particle physics, since the extra dimensions were assumed to be 
tiny, and the new particles were predicted to have masses around the Planck scale, i.e. $10^{16}$~TeV. 
Though there had been some speculations that there could be extra dimensions which were much bigger 
\cite{541,542,543}, these did not attract much attention at the time, even though a series of papers by 
I.~Antoniadis and collaborators \cite{544,545,546,547,548} went on to suggest that one or more of the 
extra dimensions in a string theory could be large and have observable effects.
          
The watershed was a 1998 paper \cite{549} by Arkani-Hamed, Dimopoulos and Dvali, now collectively 
shortened to 'ADD', in which it was suggested that the presence of large extra dimensions could eliminate 
the hierarchy problem by bringing down the Planck scale to a few TeV. To ensure that this remained 
compatible with the weak strength of gravity, it was necessary to assume that our familiar 
four-dimensional Minkowski world is actually a wafer-thin topological defect in the higher dimensional 
'bulk spacetime', on which the SM fields have somehow got trapped. Almost immediately after, the authors 
teamed up with Antoniadis to place this in the context of string theories \cite{550}. As a matter of fact, 
string theories can provide just the sort of topological defect with trapped fields as imagined by ADD --- 
these are called $D$-branes and had been invented a decade earlier by Polchinski and collaborators 
\cite{551} and by Horava \cite{552}. From this point, it became common for the fourdimensional topological 
defect to be called 'the brane', even if there was no actual use of string theory in the problem. However, 
it is likely that this model would have remained a curiosity, had not the low-energy theory with 
multiple-graviton emission and exchange been explicitly worked out by two different groups \cite{553, 
554}, later in the same year. It was these, together with a paper by Mirabelli {\it et al.} \cite{555}, 
that set off the hunt for extra dimensions at colliders, including the LHC, which was then definitely on 
the way.

Indian involvement with extra dimensions began early -- in 1998 itself 
-- with the trio of Mathews, Raychaudhuri and Sridhar \cite{556,557,558}, who 
investigated Kaluza-Klein graviton effects in $t\bar{t}$ production, 
$ep$ scattering and dijet production, in the context of the Tevatron and 
HERA colliders. This set off an explosion of studies in the Indian HEP 
community, some of which, relating to the LHC, is summarised below. As a 
matter of fact, extra dimensional theories soon split into three 
different paradigms, viz.
\vspace*{-0.2in} 
\begin{itemize} 

\item The Large Extra Dimensions (LED) paradigm, which was essentially the model of ADD, where two or more 
extra dimensions are of size as large as, perhaps, a millimetre or a few hundred microns, and the SM 
fields are confined to a single brane of thickness less than $(100~GeV)^{-1}$, while gravitons move freely 
in the toroidal bulk.

\item The Warped Extra Dimensions (RS) paradigm, proposed by Randall and Sundrum \cite{559} in 1999, in 
which there are two branes at the ends of a single extra dimension with the topology of a once-folded 
circle ($S(1)/Z_2$)and fine-tuned vacuum energies in the bulk and on the branes. This configuration 
permits a solution of the Einstein equations where the Planck mass can be small on one brane and 
exponentially large on the other.

\item The Universal Extra Dimensions (UED) paradigm, proposed by Appelquist {\it et al.} \cite{560} in 
2001, which envisages a bulk geometry similar to the RS model, but places all the SM particles in the 
bulk. The existence of a discrete symmetry due to the specific geometry acts in a way similar to 
$R$-parity and is indeed known as KK-parity. This allows this model to have a candidate for dark matter in 
the same way as SUSY does.  \end{itemize} \vspace*{-0.2in} If we go by these divisions, work on the LED 
paradigm was somewhat limited, since the other, more interesting options were soon invented. Nevertheless, 
some studies of possible LED signals at the LHC, and associated phenomena were made over the next decade 
or so \cite{561,562,563,564,565,566,567,568,569,570,571}. Studies of QCD corrections to collider processes 
in LED models have mostly been made by Mathews and Ravindran with their students and collaborators 
\cite{571,572,573,574,575,576,577,578,579,580,581,582,583,584,585,586,587,588}.

The minimal RS model, which predicts heavy graviton resonances, has a phenomenology more akin to more 
conventional electroweak models \cite{589}. Therefore, there were a few studies from India of the early 
versions of the RS model \cite{590,591,592,593,594,595,596,597,598}, including QCD corrections to the 
relevant processes \cite{598,599,600, 600,601,602,603,604}. Later came the so-called bulk RS models, where 
all the SM particles can be in the bulk (as in UED models) and have Kaluza-Klein modes on the brane. These 
came is stages, with the gauge bosons, and then the fermions, and finally the Higgs boson field, being 
moved into the bulk. These models have been mostly studied in India by Sridhar and his collaborators 
\cite{605,606,607,608,609,610,611,612,613,614,615,616,617,618,619,620,621,622,623,624}. Extensions of the 
minimal RS model have been attempted in the gravity sector by Choudhury and collaborators 
\cite{625,626,627,628,629,630,631,632}, while effects of a bulk torsion field have been discussed at 
length by Mukhopadhyaya and SenGupta in a series of papers 
\cite{633,634,635,636,637,638,639,640,641,642,643,644,645,646,647,648,649}. The latter has also studied 
the possibility of Gauss-Bonnet and higher curvature terms in the bulk 
\cite{650,651,652,653,654,655,656,657,658,659,660,661,662}.

It was also realised, soon after the RS model was proposed, that it requires an extreme fine-tuning of the 
inter-brane separation to get the right hierarchy of mass scales. To obtain this more naturally, 
Goldberger and Wise added an additional bulk scalar to the model \cite{663, 664}, which appears on the 
brane as a kind of dilatonic field, dubbed the radion. The phenomenology of a light radion is quite 
similar to that of the Higgs boson, and in fact, it can mix with the Higgs boson \cite{665}. Partly for 
this reason, there has been considerable interest in the Indian community on radion physics and signals 
for radions at colliders, including the LHC \cite{666,667,668,669,670,671,672,673,674,675}.

The minimal UED models (mUED) have also been studied quite intensively in India \cite{189, 370, 371, 
676,677,678,679,680,681,682,683,684,685,686,687,688}. Attempts to extend this to non-minimal versions have 
been made in two directions --- one by assuming more than one universal extra dimension 
\cite{689,690,691,692,693,694,695} and one by assuming boundary-localised terms in the action 
\cite{696,697,698,699,700,701,702,703}. In either case, changes happen in the mass spectrum and the 
couplings and these reflect in the collider phenomenology.
       
Two offshoots of theories with extra dimensions bear mentioning. One was the rather arcane idea of 
unparticles \cite{704,705,706} --- which are field theoretic configurations in four dimensions with a 
behaviour similar to higher dimensional fields and a phenomenology rather similar to that of the ADD model 
(2007). This idea inspired quite a few Indian phenomenological studies of unparticles 
\cite{707,708,709,710,711,712,713,714,715,716,717,718,719,720,721,722,723,724,725,726,727} before the idea 
quietly faded out of the limelight. Somewhat more lasting were an ingenious class of models, collectively 
known as 'little Higgs models' \cite{728,729,730,731}. In these models, developed between 2000 and 2005, a 
gauge group $[SU(2) \times U(1)]^n$ breaks in stages to the $SU(2)_L \times U(1)_Y$ of the SM, creating a 
hierarchy of heavy gauge bosons, and heavy fermions to match. The beauty of this model -- or class of 
models -- lies in the fact that group-theoretic factors provide the negative signs required for 
cancellation of quadratic divergences in the Higgs boson self-energy, which can, therefore, happen 
separately \cite{732} between pairs of bosons and pairs of fermions\footnote{Unlike SUSY, where these 
happen between boson and fermion pairs.}. In some versions of these models, the extra fields have a $_2$ 
symmetry called $T$-parity, as a result of which the lightest particle with non-zero $T$ becomes stable 
\cite{733,734,735,736}. This model, therefore, has a solution to the hierarchy problem together with a 
dark matter candidate, just as SUSY has. Indian HEP theorists were somewhat late getting on to this 
bandwagon, but eventually this model, or rather class of models, did inspire a whole series of 
investigations \cite{161, 307, 337, 
737,738,739,740,741,742,743,744,745,746,747,748,749,750,751,752,753,754,755,756,757}, before they, too, 
faded into the twilight with the others.

\section{Continuing with caution}\label{sec6}

Gradually, as the LHC continued to run beyond 2012, the euphoria accompanying the Higgs boson discovery 
began to fade. As we have noted before, there have been no signs in the intervening decade of any signals 
for compositeness, SUSY, extra dimensions, or any kind of BSM physics whatsoever \cite{210}. So uniformly 
have the data matched with the predictions of the SM, that the LHC is already being dubbed 'the Standard 
Model collider'. Moreover, the entire paradigm of hypothesis and prediction preceding discovery has been 
called -- perhaps prematurely -- into question \cite{212}. And yet, theorists are not comfortable with a 
world where the SM is the ultimate theory of elementary particles and their interactions. It is useful, at 
this juncture, to briefly review the reasons for this profound dissatisfaction.

Apart from discomfort with the large number of experimentally fitted parameters in the SM, which seem to 
follow no known order or system, theorists have seven major issues with the SM as a theory. These can be 
classified into three groups, viz.
\vspace*{-0.2in}
\begin{itemize}

\item {\sf Missing components}: The biggest puzzle is the nature of (I) {\it dark matter} and (II) {\it 
dark energy}. If these are particulate in nature -- as seems to be the general consensus, for dark matter 
at least -- then there are no candidates in the SM which fit the bill.

\item {\sf Inadequate parameters}: One of these issues is (III) {\it baryogenesis}, or the 
matter-antimatter asymmetry of the Universe, which may be tracked to the measured level of $CP$-violation 
in the SM being some $10^{-11}$ times too small to explain the observed baryon asymmetry. On the other 
hand, we also have the (IV) {\it strong $CP$ problem}, where the $\theta_{\rm QCD}$ coupling has to be 
tuned to less than $10^{-10}$ in the SM without apparent justification. There is also the issue of (V) 
{\it neutrino masses}, which were originally believed to vanish, but have been proved otherwise. 
Accommodating them in the SM leads to multiple issues of naturalness and lepton number non-conservation.

\item {\sf Inconsistency at quantum level}: It is beyond question that the SM is a quantum field theory, 
for this has been vindicated by a large body of evidence, including neutral pion decay, precision 
measurements of electroweak parameters, the running of the strong coupling constant and the occurrence of 
rare flavour-changing decays. However, as described in Sect.~1, this leads to a (VI) {\it hierarchy 
problem}, with a fine-tuning of the running Higgs boson self-energy to some 16 orders of magnitude. 
Another problem arises with the running of the Higgs boson self-coupling, which tends to destroy (VII) 
{\it vacuum stability} at some high-energy scale, something that can only be countered by invoking BSM 
physics.

\end{itemize} 
\vspace*{-0.2in} 

Earlier UV-complete theories like SUSY provided simultaneous answers to several of these questions at the 
same time. Unfortunately, the whole paradigm of UV-complete theories predicting light particles 
discoverable as resonances at the LHC -- though by no means disproved -- has lost popularity as the best 
way to probe BSM physics at the LHC. Instead we have about half-a-dozen new approaches, which eschew the 
top-down approach, and instead, concentrate on minimal extensions, effective field theories, simplified 
models, and so on, i.e. essentially bottom-up approaches which do not worry about issues of UV 
completeness or nonunitarity. The high-energy physics community has certainly learnt the virtues of 
caution! In this section, we briefly discuss a few of these approaches, focussing on the Indian 
contributions to this field and restricting ourselves to the LHC context. Since this is an active area of 
research, with new ideas being proposed on an almost daily basis, this section will necessarily be 
incomplete, even sketchy.

\subsection{Minimal extensions}\label{subsec6.1}

The SM has three sectors, viz. the gauge boson sector, the fermion sector and the scalar sector. Minimal 
extensions have been contemplated in all of these. The oldest and perhaps the best-motivated of these is 
the left-right symmetric model (and its SUSY version), based on a group $SU(2)_L\times SU(2)_R\times 
U(1)_{B-L}$ which breaks spontaneously to $SU(2)_L \times U(1)_Y$, thereby explaining the origin of parity 
violation \cite{64,65,66,67,68,69}. This class of models predicts the existence of heavy $W_R$ bosons, and 
there is a considerable body of work by Indians discussing collider signatures for these or closely 
related phenomena 
\cite{758,759,760,761,762,763,764,765,766,767,768,769,770,771,772,773,774,775,776,777,778,779,780}. Some 
gauge extensions, as well as a few recent 'anomalies' in the flavour sector seem to call for the existence 
of an extra U(1) symmetry and this predicts a heavy $Z'$ boson, which has also been studied quite 
intensively by Indian theorists \cite{781,782,783,784,785,786,787}.

In the fermion sector, the LEP precision measurements of electroweak parameters \cite{788} permit the 
existence of just three kinds of heavy BSM fermions, viz. sequential fermions with left-handed doublets 
and right-handed singlets, mirror fermions with left-handed singlets and right-handed doublets, and also 
vectorlike fermions, where both left- and right-handed fermions are either singlets or doublets. A fourth 
sequential generation, however, was ruled out by the LEP data except for a highly fine-tuned version 
\cite{789,790,791,792,793}, and even that was ruled out by the LHC data soon after the Higgs boson 
discovery \cite{794,795,796,797}. Mirror fermions, however, and vectorlike fermions, survive to this day, 
and, together with other classes of exotic fermions, have been investigated by several Indian theorists 
\cite{776, 798,799,800,801,802,803,804,805,806,807,808,809,810,811,812,813,814,815,816,817,818,819,820}, 
most notably Gopalakrishna \cite{57, 821,822,823,824,825,826} and his collaborators. Another class of 
exotic fermions are the bulk fermions in a higher dimensional theory, which have already been mentioned in 
the previous section.

Finally, we come to the Higgs sector. Since we know least about this sector, except that a Higgs boson 
exists, we have the greatest freedom to propose modifications in this part of the SM. Many of these have 
been discussed in Sect.~3. In addition, a doubly charged Higgs boson, such as arises whenever there is a 
scalar triplet or bidoublet in the model, has received considerable attention 
\cite{327,827,828,829,830,831,832,833,834} in the Indian community.

\subsection{Modelling Dark Matter}\label{subsec6.2}

There is a wide consensus today that the dark matter component of the Universe is particulate in nature, 
and that these are BSM particles having gravitational and weak interactions, but not strong and 
electromagnetic interactions. The gravitational interactions have been proved from astrophysical 
observations, and weak interactions are necessary for enough dark matter to be produced to satisfy the 
relic density inferred from CMB studies. The absence of electromagnetic interactions makes them 'dark' and 
the absence of strong interactions makes them 'non-baryonic', both of which are proved from astrophysical 
observations \cite{835,836,837,838}.

There has been much speculation about these particles. The Indian contribution to this has mostly been 
along two lines, viz. 'invisible' particles and 'simplified models'. In the first paradigm, the existence 
of weakly interacting massive particles (WIMPs) has been assumed and the analyses discuss how to search 
for and/or constrain such theories \cite{172, 396,569,839,840,841}. In the second paradigm, the SM is 
minimally extended to include a dark matter particle, and one or two 'mediators' to ensure that the dark 
matter has weak interactions \cite{841,842,843,844,845,846,847,848,849,850,851,852,853,854,854}. Until 
there is some experimental evidence from the LHC or from direct search experiments, the field is wide open 
for speculation, somewhat as the whole of BSM physics was in the 1980s.

\subsection{Compressed spectra}\label{subsec6.3}

When heavy particles are produced at high-energy colliders, they immediately decay into lighter particles, 
which in turn, decay into lighter particles, and the process continues until there are stable particles in 
the final state. Such a cascade decay process creates a 'signature' of the final particle in the form of a 
pattern of stable co-produced final states \cite{877,878,879}. This has always been the classic technique 
to look for new particles in the SM, and now beyond it, in models such as SUSY and extra dimensions. 
However, if the mass gap between the initial and final states is small, i.e. the mass spectrum of all the 
intermediate states are 'compressed' into a small range, then the (meta)stable particles in the final 
state, viz. electrons, photons, muons, etc. will have very low energies and may fall below the triggering 
thresholds of the detectors. As a result, in the case of compressed mass spectra, a BSM model becomes very 
difficult to discover. SUSY models, in particular, have some parts of the allowed parameter space which 
predict such spectra. It has, thus, been a part of the international effort, to which Indians have also 
contributed \cite{315,316, 390,395,880,881}, to identify novel variables and signatures which would enable 
us to find new particles, even in the case of such compressed spectra.

\subsection{New variables for BSM studies}\label{subsec6.4}

Inspiration from the study of compressed spectra has lead researchers to explore new kinematic variables 
where signatures for new physics may be lurking and to devise ways and means of exploiting the existent 
data to the maximum using such variables \cite{337,855,856,857,858,859,860,861,862,863,864,865,866,867, 
868,869,870,871}. Inevitably, such searches have led to the use of machine learning tools to sift and find 
patterns in the data. This field has been attracting an increasing amount of interest from India 
\cite{327, 872,873,874,875,876} in recent times and holds out the promise of being a major tool in the 
future runs of the LHC.

\subsection{Effective Field Theories}\label{subsec6.5}

In this class of models, the SM is assumed to be the low-energy limit of a deeper, presumably UV-complete, 
theory at a higher energy scale. The prototype of this is the Fermi effective theory of weak interactions, 
which is the low-energy limit of the gauge theory of Glashow, Salam and Weinberg. Just as the Fermi 
current-current operator has mass dimension 6 and the coupling constant $G_F$ has a negative mass 
dimension, so do these effective field theories, or EFTs, involve higher dimensional operators with 
dimensional couplings, presumably reflecting the scale of the new physics. The total number of such 
operators at dimensions 5, 6 and 8 (which have mostly been studied) is very large \cite{882}, but there 
are gauge-invariant subsets under renormalisation group evolution, which makes the problem more tractable 
\cite{883}. Moreover, for a particular set of processes, the dominant contributions may come from a 
limited set of operators, leading to an EFT version of simplified models. Many of these aspects have been 
investigated by Indian physicists \cite{884,885,886,887,888, 889,890,891,892,893,894,895,896,897,898,899}, 
though much remains to be done, as this is an active, ongoing line of research.

\section{Pontifications \& predictions}\label{sec7}

Having reviewed the happenings in this field during the past 40-odd years, one can essay to make some 
remarks of a general nature. Perhaps the first and most important observation to make is that the element 
of uncertainty in high-energy physics (HEP), is perhaps, the highest in any current branch of science, 
since what is being probed is a length scale never 'seen' before and, if our current understanding of a 
Big Bang origin of the Universe is correct, a time scale in the early Universe which has never before been 
'accessible'. The history of science in the nineteenth and twentieth centuries is replete with examples of 
the big surprises which awaited researchers when they probed smaller scales. Of these, the discovery of 
atomic sub-structures and the composite nature of hadrons were perhaps the most dramatic, but of equal 
importance was the discovery that it is continuous fields, and not corpuscular entities, that form the key 
to understanding Nature at such small lengths. The importance of symmetry as a framework to develop 
fundamental theories has also been gradually forced upon us, until it has now become the first tool of the 
HEP theorist. And yet we do not really know what lies at length scales below $10^{-17}$~cm, and most of 
our guesses do not seem to be working. High-energy physicists are, therefore, once again as they were in 
the first half of the twentieth century, like pioneers pushing into an unknown world, unaware of what 
exotic landforms or animals may be found there.

During the second half of the twentieth century, and indeed, until a decade ago, HEP witnessed a very 
different paradigm compared to the rest of the scientific world. In HEP, theoretical ideas surged ahead of 
experimentation, with new theories making predictions far beyond the capacity of contemporary experiments 
to test. Of these, the set of ideas clubbed together as the Standard Model, have proved successful far 
beyond the dreams of anyone who proposed it. Today, in the sheer number of experimental results which 
match with the Standard Model, arguably the only competitor would be Newtonian gravity.

However, this success has been very hard-earned. From the 1970s, as high-energy physics moved away from 
cosmic ray and fixed target to colliding-beam experiments, the complexity and cost of high-energy physics 
experiments has shot up by many orders of magnitude. Modern HEP experiments -- especially of the collider 
type -- have become notorious for their size, complexity, long gestation time --- and, therefore, cost. 
The size and scope of these experiments requires a small army of research workers, from engineers to data 
analysts, to run the machines, extract the data and interpret them. Given that the quantity of data to 
process increases rapidly with the energy, this makes for a long time before we can expect results from a 
new experiment. The involvement of more people also increases the possibility of human errors, and thus 
further delays are introduced while the results go through multiple checks before obtaining a stamp of 
approval from the relevant collaboration. Thus, it took a full half century as the Standard Model and its 
different predictions were verified, particle by particle and process by process, at different 
experiments. All the while, theorists were left free to let their minds range over the smaller length 
scales which would one day become accessible to experiment, and make predictions for the same.

\subsection{Modern Times}\label{subsec7.1}

In the post-Higgs discovery era -- essentially the past decade -- we have seen the runaway success of 
theoretical ideas, especially those based on new symmetries, come to a grinding halt. The LHC has been 
slowly, but surely, checking the predictions of these exotic theories, and, till date, has come up with a 
total blank. However, the LHC has barely scratched the surface of the new energy regime, and even if it 
goes deeper with the same kind of results, all that this would mean is that physics at the new length 
scale must be different from what has been imagined -- or can be imagined -- and it will require 
experimental data to create a new understanding of it. That is, and always has been, the sober scientific 
view. However, there are always sociological ripples, including detractors who would argue that 
mathematical approaches to physics are fundamentally wrong, or that too much tax-payers' money is being 
spent, or that scientists should not dare to speculate beyond a certain limit. Generally, in science as in 
any other field of human endeavour, such voices rise during a period of stagnation and fall silent 
whenever the next major advance/breakthrough comes.

In truth, as the saying goes, rumours of the death of BSM physics have been greatly exaggerated. What has 
mostly been eliminated is the most 'hopeful' part of the vast parameter space of these models, where the 
word 'hopeful' is used in the sense of that which would have given new physics in the early runs of the 
LHC. However, Nature has not proved to be so generous, and in fact, rarely is. In the 1970s, the $W$ and 
$Z$ bosons were expected to lie around 2~GeV and in the 1970s, the Higgs boson and the top quark were 
believed to lie around $10-20$~GeV. These were the 'hopeful' values at the time, but it required much 
higher energies and much higher costs to actually find these particles. Nevertheless, they do exist and 
the entire human race would have been the poorer if their search had been abandoned in the 1980s. Thus, 
all that we can say is that there is a long road ahead, where new physics of the expected, deviant or 
totally unexpected kind may appear at any juncture.

Another point to note is that the style of theoretical work in HEP has been evolving quite rapidly over 
the past two decades. Till the end of the twentieth century, theorists contented themselves with 
calculating total and differential cross-sections, and making fairly crude numerical estimates using some 
broad kinematic cuts to simulate the experimental constraints. It was left to the experimentalists to 
simulate the events, including all peripheral effects and the limitations of their own detectors. However, 
mostly after the turn of the century, there has been a proliferation of 'user-friendly' software packages, 
with different levels of sophistication, to be used in the analyses by theorists. Thus, we have packages 
like, to name only a few, MadGraph, Pythia and Herwig to simulate collider processes, SuperISO and EOS for 
flavour physics, MicrOMEGAs to calculate the dark energy relic density and DELPHES to simulate detector 
effects. This is accompanied by mathematical calculation tools like Form, FeynRules and, of course, the 
all-pervading Mathematica. Recently, as mentioned in Sect.~6.4 above, the use of Machine Learning tools 
has also come into vogue, and no doubt the near future will see these incorporated into some of the HEP 
packages.

Thus, a young theorist now requires to master a large number of software tools to remain competitive, but 
once this is done, the calculational process is accelerated, leaving time for the researcher to think 
about deeper issues. It may thus be hoped that with these powerful tools, the next generation of theorists 
will be able to make inroads into problems which have eluded their predecessors. On the flip side is the 
caveat that the power and beauty of these software tools should not intoxicate researchers into creating 
unnecessary sophistication where it does not add any insight\footnote{In this context, readers may 
appreciate Richard Feynman's observations regarding his colleague Stan Frenkel, in Ref. \cite{900}.}.

\subsection{The Indian context}\label{subsec7.2}

Turning now to the Indian context, HEP theory and experiment have grown greatly over the past 4 decades. 
For the experimental effort, the 1970s and 1980s saw a drastic decline in numbers, as cosmic ray 
experiments receded into the background, and accelerator physics grew in proportion. The closure of the 
Kolar gold mines in 1991 led to the abandonment of India's front-line proton decay experiment. Henceforth, 
Indian experimental participation in cutting-edge HEP would be through international collaborations --- at 
CERN, FNAL and KEK, among others. The first groups to get into this were from TIFR and BARC and the 
Universities of Delhi and Punjab, but others soon joined in. Many of the graduate students in the early 
years moved abroad, but increasingly many are coming back to India after postdoctoral stints. The growth 
in the number of science-oriented institutions in the country has also helped many of these young HEP 
experimentalists to find positions and set up laboratories under the aegis of one international 
collaboration, or another. The pool of experience in India has steadily grown till it now includes seniors 
who have worked on the ISR and fixed target experiments at CERN and at the proton decay experiment at 
Kolar, and a large number of others who have worked on the collider experiments at SLAC and FNAL in the 
USA, the LEP and LHC machines at CERN, as well as the BELLE experiment at KEK. Sadly, India's own attempt 
to set up a large HEP experiment in the form of an underground neutrino detector to be called the 
India-based Neutrino Observatory (INO) has remained bogged down in socio-political issues till the present 
juncture.

On the theory side, too, there has been a massive growth in the number of researchers. An interesting 
phenomenon, indicative of a mature community, is that there are now four generations of HEP theorists who 
are active in the country. The author notes that this includes his own teachers and mentors, his own 
colleagues and contemporaries, as well as his students and their students. This is a healthy sign, no less 
than the fact that the present HEP community is spread over a wide variety of academic institutions across 
the country and there exist groups where earlier there used to be a solitary practitioner. Another sign of 
maturity is the fact that there are now many close collaborations between theorists and experimentalists, 
something which would have been very rare in the 1980s. There is an increasing trend for close 
collaborations between particle physicists and astrophysicists, as well as cosmologists, as the two 
subjects show signs of merging, especially in the area of Dark Matter. On the other hand, HEP theory and 
string theory have moved away from the close synergy they had in the past century, partly because string 
theory is moving closer to condensed matter physics, but also because supersymmetry, the main tie between 
the two areas, is getting pushed into the background.

The new HEP theorists in India have, in general, embraced the software-dominated culture of HEP theory 
very well. Fortunately most of these softwares lie in the public domain, and those that do not are quite 
reasonably priced. As a result, Indian HEP theorists are not financially constrained as Indian HEP 
experimentalists are, and can acquire as much skill as their foreign counterparts in more scientifically 
advanced countries. This shows up in the high quality of theory HEP publications from India. The next step 
would be for some of the software packages to actually be developed in India. Given the success of Indians 
and Indian-origin workers in the IT industry, such a consummation may be predicted with some level of 
confidence.

\subsection{The future}\label{subsec7.3}

Now, finally, we come to the paramount question --- what results can we expect in HEP in the years to 
come? One answer is obvious, viz. we shall see the SM being tested at higher energies and luminosities, 
i.e. in effect, at a smaller length scale. It is also easy to predict that most of these tests will be 
along the lines of the past, yielding more confirmation of the SM predictions. The excitement, if any, 
will lie in anything which deviates from the SM and raises hopes that new physics has -- at last -- been 
discovered.

Over the years, there have been several false alarms and abortive signals for BSM physics. These have 
generally disappeared after the collection of more data, or a re-evaluation of the SM prediction. Among 
the few which have persisted are the muon anomalous magnetic moment $(g - 2)_\mu$ \cite{901, 902} and a 
handful of flavour anomalies \cite{903}. Most recently, the complete data from the now-defunct CDF 
experiment at Fermilab has been analysed to yield a value of the $W$ boson mass which shows as much as a 
$7\sigma$ discrepancy with the SM prediction \cite{904}. Though it has immediately become the subject of 
intense speculation, including by Indians \cite{905,906,907,908,909,910,911,912,913,914}, this result will 
need to be confirmed in the further runs of the LHC before it can be taken in all seriousness. One can 
only say that, taken together, all these anomalies constitute a tantalising set of hints that the SM may 
not be the final theory of elementary particles, after all.

The LHC is scheduled to run for another decade and a half \cite{915}, and there is already much 
speculation about the next generation of high-energy colliders. Given the pool of talents and expertise in 
the burgeoning high-energy physics community in India, we shall no doubt see a great deal more of 
LHC-oriented research and beyond coming out of India, and -- who knows? -- maybe a major discovery. It is 
an exciting thought.

\bigskip
 
\noindent{\small {\sl Acknowledgements}: The author acknowledges support of the Department of Atomic 
Energy, Government of India, under Project Identification No. RTI 4002a}.

\def\baselinestretch{1.05}

\end{document}